\documentstyle[11pt,graphicx]{article}
\title{Modifications by QCD transition and $e^+e^-$ annihilation
on analytic spectrum of relic gravitational waves
in accelerating universe}

\author{\small S. Wang$^1$, Y. Zhang$^1$\cite{email},
               T.Y. Xia$^1$, and H.X. Miao$^{1,2}$  \\
   \small $^1$Astrophysics Center,
         University of Science and Technology of China,
         Hefei, Anhui, 230026, China \\
   \small $^2$AIGO, University of Western Australia, WA 6009, Australia}
\date{}

\begin{document}
\maketitle
\baselineskip=19truept

\def\vek{\vec{k}}
\renewcommand{\arraystretch}{1.5}
\newcommand{\be}{\begin{equation}}
\newcommand{\ee}{\end{equation}}
\newcommand{\ba}{\begin{eqnarray}}
\newcommand{\ea}{\end{eqnarray}}

\sf

\begin{center}
\Large  Abstract
\end{center}

\baselineskip=19truept

\begin{quote}

As predicted by quantum chromodynamics(QCD),
around $T\sim 190$ MeV in the early universe,
the QCD transition occurs during which the quarks are
combined into the massive hadrons.
This process reduces the effective relativistic degree of freedom,
and causes a change in the expansion behavior of the universe.
Similarly,  the $e^+e^-$ annihilation occurred around $T\sim 0.5$ Mev
has the same kind of effect.
Besides, the dark energy also
drives the present stage accelerating expansion.
We study these combined effects on the relic gravitational waves (RGWs).
In our treatment, the QCD transition and the $e^+e^-$ annihilation,
each is respectively represented by a short period
of expansion inserted into the radiation era.
Incorporating these effects,
the equation of RGWs is analytically
solved for a spatially flat universe,
evolving from the inflation up to the current acceleration,
and the spectrum of RGWs is obtained,
covering the whole range of frequency $>10^{-19}$ Hz.
It is found that the QCD transition
causes a reduction of the amplitude of RGWs by
$\sim 20\%$ in the range $>10^{-9} $ Hz,
and the $e^+e^-$ annihilation causes a reduction $\sim 10\%$
in the range $>10^{-12} $ Hz.
In the presence of the dark energy,
the combination of the QCD transition and the $e^+e^-$ annihilation,
causes a larger reduction of the amplitude
by $\sim  30\% $ for the range  $>10^{-9} $ Hz,
which covers the bands of operation of LIGO and LISA.
By analysis, it is shown
that RGWs will be difficult to detect by the present LIGO,
but can be tested by LISA for certain inflationary models.

\end{quote}


\newpage

\begin{center}
{\bf 1. Introduction}
\end{center}

The existence of a stochastic
background of relic gravitational waves (RGWs)
is generally predicted in inflationary models
\cite{grishchuk,starobinsky,zhang2}.
Since the relic gravitons decoupled much earlier
than the cosmic microwave background (CMB) photons,
the detections of RGWs would open a new window to
the very early universe.
Unlike gravitational radiations from
finite stellar objects,
RGWs exist everywhere and anytime,
and have a wide spreading spectrum,
serving as one of the major scientific goals
of the laser interferometers gravity-wave detections,
including the current LIGO \cite{ligo}, VIRGO \cite{virgo},
GEO600 \cite{geo}, TAMA \cite{tamma}, AIGO \cite{aigo},
and the future LISA \cite{lisa}, ASTROD \cite{astrod},
BBO \cite{bbo}, and DECIGO \cite{decigo}.
Moreover, along with the density perturbations,
RGWs also contribute to the  CMB anisotropies and polarizations
\cite{basko,Kamionkowski,pritchard, zhao2, baskaran}.
For instance,
the B-polarization of CMB on very large scales
can only be generated by RGWs.
Through the detections of CMB polarizations
one may have a chance
to obtain the direct evidence of GWs for the first time
\cite{ Spergel,Tegmark,seljak1,page,hinshaw}.
Therefore, the precise information of RGWs is much desired.

The spectrum of RGWs  depends on the
following factors.
First, it depends sensitively on the specific
inflationary models \cite{grishchuk,zhang2,zhao}.
After being
generated, it will be altered by the subsequent stages of expansion
of the universe, among which notably is the currently accelerating
expansion of the universe \cite{zhang2, perlmutter}.
Finally, it
would be further modified by other physical processes occurred in
the early universe.
For example, the neutrino free-streaming
\cite{weinberg} has been shown to affect the spectrum of RGWs in the
very low frequency range $(10^{-16}\sim 10^{-10})$ Hz \cite{Miao}.
Although this would modify the contribution to CMB polarizations,
but would not affect the detections of LIGO \cite{ligo} and LISA
\cite{lisa} that operate effectively at higher frequencies,
$(10^{1}\sim 10^{3})$ Hz and $(10^{-7}\sim 10^{0})$ Hz, respectively.
Other important physical
processes include the QCD transition
around a temperature $T\sim 190$ MeV \cite{cheng,karsch},
and the $e^+e^-$ annihilation around $T\sim 0.5$ MeV.
These two processes will be studied in this work.
Leaving aside the
detail of the QCD transition, which is notoriously complicated and
still under study, we are concerned with only the thermodynamic
property that the transition brings to the system, i.e., a
significant drop in the relativistic degrees of freedom.
Consequently, the constituent components in the stress tensor
$T_{\mu\nu}$ also change in the Friedmann's equation.
Thereby the
expansion of the Universe in terms of the scale factor $a(\tau)$
will subsequently change, and the spectrum of RGWs would be
modified.
The $e^+e^-$ annihilation has the similar effect.
Schwarz \cite{schwarz} studied
the QCD transition and the $e^+e^-$ annihilation
and estimated the reduction of the energy density spectrum of RGWs.
Watanabe and Komatsu \cite{yuki} investigated these effects with more
details, and gave a numerical solution of the energy density
spectrum of RGWs.
In these works, the important effect of the
accelerating expansion of the present universe \cite{zhang2} has not
been included, neither the effects of inflation and reheating.
Moreover, to show the prospects of detecting RGWs
at the laser interferometers gravity-wave detection,
an explicit demonstration of the current
spectrum of RGWs itself and its direct comparison
with the sensitivity curve of gravity-wave detection
are also needed.

In this paper,
by extending our previous analytical calculation of the spectrum of RGWs,
we explore the consequences caused by the QCD transition
and the $e^+e^-$ annihilation,
and, at the same time, take into account
of the accelerating expansion driven by the dark energy  \cite{zhang2}.
Beside the energy density spectrum $\Omega_g(\nu)$,
we demonstrate explicitly the spectrum $h(\nu, \tau_H)$ of RGWs itself,
and compare it directly with
the sensitivity of the ongoing and forthcoming GW detectors,
such as LIGO, LISA, etc \cite{ligo,lisa}.
Aiming at giving a comprehensive compilation,
by using the set of parameters
$\beta$, $\beta_s$, $n$, $v$, $\gamma$ and $r$, respectively, such
important cosmological elements have been explicitly parameterized,
as the inflation, the reheating, the QCD transition,
 the $e^+e^-$ annihilation,
the dark energy and the tensor/scalar ratio.
This will considerably facilitate
further studies on the RGWs and the relevant physical processes.
For instance, it can be easily used in
the calculation of CMB anisotropies and
polarizations generated by RGWs \cite{zhao2,page}.

The organization of this paper is as follows.
In section 2, from the inflation up to the acceleration,
the scale factor $a(\tau)$ is specified by
the continuity conditions for the subsequential stages of expansion.
The periods of the QCD transition and the $e^+e^-$ annihilation
are also modelled as
having a scale factor $a(\tau)$ of power-law form,
which is inserted into the radiation era.
In section 3, the analytical solution of RGWs
in terms of Bessel's functions is determined with the
coefficients being fixed by continuity condition joining two
consecutive expansion stages.
The impact of the dark energy on the expansion
is emphasized.
In section 4, we present the
resulting spectra of RGWs and discuss the modifications
caused by the QCD transition, the $e^+e^-$ annihilation,
and other effects.
Appendix A supplies the details
of our treatment of the period of QCD transition and
$e^+e^-$ annihilation by modelling the corresponding scale factor.
Appendix B gives  an interpretation of
the modifications on RGWs due to the QCD transition.
In this paper the unit with
$c=\hbar=k_B=1$ is used.

\

\begin{center}
{\bf 2. Expansion history of the universe}
\end{center}

From the inflationary up to the currently accelerating stage,
the
expansion of the universe can be described by the
spatially flat ($\Omega_\Lambda +\Omega_m+\Omega_r=1$)
Robertson-Walker spacetime with a metric
\be
ds^2=a^2(\tau)[-d\tau^2+\delta_{ij}dx^idx^j],
\ee
 where $\tau$ is the conformal time.
The scale factor $a(\tau)$
for the successive stages can be approximately
described by the following forms
\cite{grishchuk2000}:

The inflationary stage:
\be \label{inflation}
a(\tau)=l_0|\tau|^{1+\beta},\,\,\,\,-\infty<\tau\leq \tau_1,
\ee
where $1+\beta<0$, and $\tau_1<0$.
This generic form of scale factor is
a simple modelling of inflationary expansion,
and the index $ \beta$ is a parameter.
The special case of $\beta=-2$ is
the de Sitter expansion of inflation.
If the inflationary expansion is driven by a scalar field,
then the index $\beta$ is related to
the so-called slow-roll parameters,
$\eta$ and $\epsilon$ \cite{Liddle},
as $\beta=-2+(\eta-3\epsilon)$ .
In this class of scalar inflationary models
one usually has  $\beta\le -2$.
Besides, the observational results of  WMAP also indicate
that $\beta$ should be slightly smaller than $-2$ \cite{Spergel}.
But, for demonstration purpose,
we allow the parameter $\beta$ to take values $>-2$
to examine the possibility of detection by GWS detectors.

The reheating stage:
\be \label{reheating}
a(\tau)=a_z|\tau-\tau_p|^{1+\beta_s},\,\,\,\,\tau_1\leq \tau\leq
\tau_s,
\ee
where
$\tau_s$ is the beginning of radiation era and $\tau_p<\tau_1$.
We will mostly take the  model parameter
$\beta_s= -0.3$, though other values are also taken
to demonstrate the effect of various reheating models.

\begin{figure}
\centerline{\includegraphics[width=8cm]{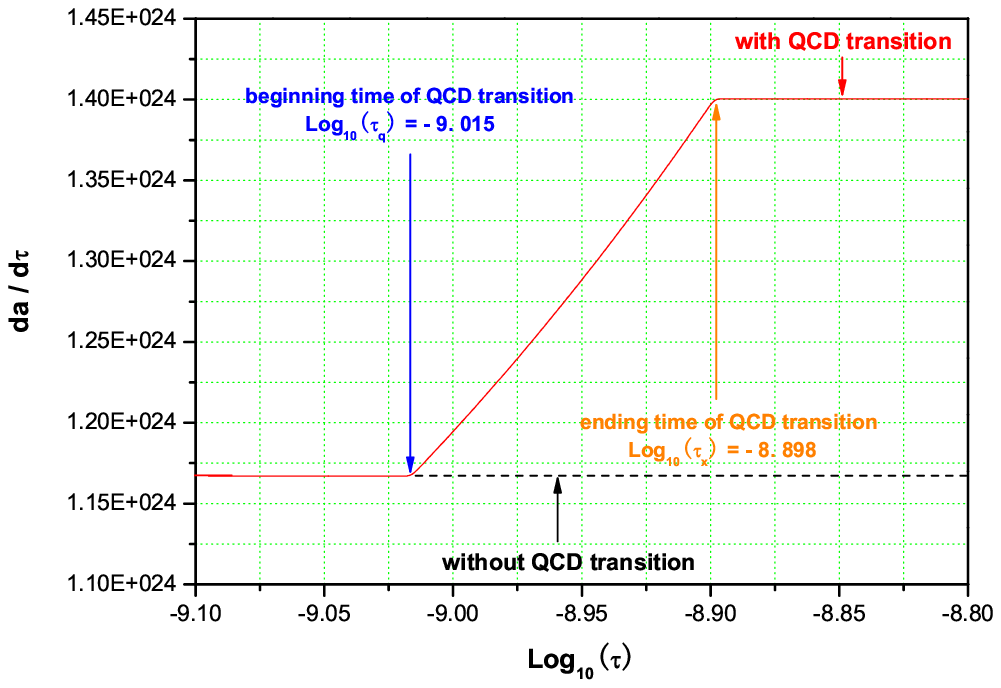}}
\caption{\label{fig1}
The derivative  $ a'(\tau)$ jumps up
around the QCD transition modelled by Eq.(\ref{rad2}).
If there were no QCD transition ,
$ a'(\tau)$ would be a constant in this graph.}
\end{figure}

The radiation-dominant stage:

The QCD transition occurs around $T\sim 190$ Mev for a period,
and the process of $e^{+}e^{-}$ annihilation
into photons  starts around $T\sim 0.5$ MeV
and ends up around $T\sim 0.1$ MeV \cite{yuki}.
These two periods  should be included into the radiation stage.
Before the QCD transition, one has
\be \label{rad1}
 a(\tau)=a_e(\tau-\tau_e),\,\,\,\,\tau_s\leq
\tau\leq \tau_q;
\ee
where $\tau _q$ is the beginning of the QCD transition and
$\tau_e<\tau_s$.
During  the  QCD transition,
$a(\tau)$ is modelled by
\be \label{rad2}
a(\tau)=a_n(\tau-\tau_n)^{1+n},
   \,\,\,\,\tau_q\leq \tau\leq \tau_x ,
\ee
where $\tau_x$ is the ending of the QCD transition
and $\tau_n<\tau_q$.
The power index  $n$ in Eq.(\ref{rad2}) is a model parameter
describing the QCD transition.
We have found that the spectrum of RGWs does not vary considerably
for the values of  $n$ in the interval $(0.7\sim 4.6)$.
For concreteness we take $n=1.634$ in calculations.
(For more details, see Appendix A).
The expansion rate $ a'(\tau)$
around $T\sim 190$ Mev is plotted in Fig.\ref{fig1},
showing a jump-up caused by the QCD transition.
After the QCD transition and
before the $e^{+}e^{-}$ annihilation, one has
\be \label{rad3}
a(\tau)=a_f(\tau-\tau_f),\,\,\,\,\tau_x\leq \tau\leq \tau_y,
\ee
where $\tau _y$ is the beginning of the $e^{+}e^{-}$ annihilation
and $\tau_f<\tau_x$.
The two slopes
$a_e$ in Eq.(\ref{rad1}) and $a_f$ in Eq.(\ref{rad3}) are related
as $a_f\simeq1.2 a_e$ by considerations of
the details of the QCD transition (See Appendix A).
Similarly,
the period of $e^{+}e^{-}$ annihilation is modelled by
\be \label{rad4}
a(\tau)=a_v(\tau-\tau_v)^{1+v},
   \,\,\,\,\tau_y\leq \tau\leq \tau_z ,
\ee
where $\tau_z$ is the ending of the $e^{+}e^{-}$ annihilation
and $\tau_v<\tau_y$.
The power index  $v$ in Eq.(\ref{rad4}),  as a model parameter,
can be taken as $v=0.063$ (see Appendix A).
After the $e^{+}e^{-}$ annihilation,
one has
\be \label{rad5}
a(\tau)=a_g(\tau-\tau_g),\,\,\,\,\tau_z\leq \tau\leq \tau_2,
\ee
where $\tau_g<\tau_z$, and
$\tau_2$ is the beginning of the matter era,
which can be taken at a redshift  $z\simeq 3454$ \cite{Spergel}.
The two slopes $a_f$ in Eq.(\ref{rad3})
and $a_g$ in Eq.(\ref{rad5})
satisfy the relation $a_g\simeq1.1 a_f$ (See Appendix A).

The matter-dominant stage:
\be \label{m}
a(\tau)=a_m(\tau-\tau_m)^2,\,\,\,\,
\tau_2 \leq \tau\leq \tau_E,
\ee
where $\tau_m<\tau_2$ and
$\tau_E$ is the beginning of the acceleration era.

The accelerating stage:
\be \label{accel}
a(\tau)=l_H|\tau-\tau_a|^{-\gamma},\,\,\,\,\tau_E \leq \tau\leq
\tau_H,
\ee
where $\tau_H$ is the present time and
$\tau_H<\tau_a$.
The index $\gamma$ in Eq.(\ref{accel})
depends on the dark energy $\Omega_\Lambda$.
By fitting with the numerical solution of
the Friedmann equation \cite{zhang2, Miao},
\be \label{Friedmann}
\Big(\frac{a'}{a^2}\Big)^2=\frac{8\pi G}{3} (\rho_\Lambda +\rho_m
+\rho_r),
\ee
where $a'\equiv da/d\tau$,
one can take
$\gamma\simeq 1.05$ for $\Omega_{\Lambda}=0.7 $,
and $\gamma\simeq 1.044$ for $\Omega_{\Lambda}=0.75 $.
The redshift of the start of this stage
depends on the specific models of the dark energy.
For instance, in the  cosmological constant model with
$\Omega_\Lambda =0.72$ and $\Omega_m =0.28$,
it starts at $z\simeq 0.37$,
and, in the quantum effective  Yang-Mills condensate
dark energy model,
it starts at $z\simeq 0.5$ \cite{Zhang}.

In the above specifications of $a(\tau)$,
there are nine instances of time,
$\tau_1$, $\tau_s$, $\tau_q$, $\tau_x$, $\tau_y$, $\tau_z$,
$\tau_2$, $\tau_E$, and $\tau_H$,
which separate the different stages.
Eight of them are
determined by how much $a(\tau)$ increases over each stage based on
the cosmological considerations.
We take the following
specifications:
$\zeta_1\equiv\frac{a(\tau_s)}{a(\tau_1)}=300$
for the reheating stage,
$\zeta_s\equiv\frac{a(\tau_q)}{a(\tau_s)}=2.895\times10^{16}$
for the first radiation stage,
$\zeta_q\equiv\frac{a(\tau_x)}{a(\tau_q)}=1.34$
for the second radiation stage,
$\zeta_x\equiv\frac{a(\tau_y)}{a(\tau_x)}=283.582$
for the third radiation stage,
$\zeta_y\equiv\frac{a(\tau_z)}{a(\tau_y)}=5$
for the fourth radiation stage,
$\zeta_z\equiv\frac{a(\tau_2)}{a(\tau_z)}=1.818\times10^{4}$
for the fifth radiation stage (see Appendix A),
$\zeta_2\equiv\frac{a(\tau_E)}{a(\tau_2)}= 3454\zeta_E^{-1}$
for the matter stage, and
$\zeta_E\equiv\frac{a(\tau_H)}{a(\tau_E)}
   =(\frac{\Omega_\Lambda}{\Omega_m})^{1/3}$
for the present accelerating stage \cite{zhang2, Miao}.
The remaining time instance is fixed by an
overall normalization
\be \label{norm}
|\tau_H-\tau_a|=1.
\ee
There are also 22 constants in the expressions of
$a(\tau)$,
among which $\beta$, $\beta_s$, $n$, $v$ and $\gamma$ are
imposed as the model parameters
describing the inflation, the reheating,
the QCD transition, the $e^+e^-$ annihilation and the acceleration,
respectively.
Based on the definition of
the expansion rate $H_0=\frac{a'}{a^2}|_{\tau_H}$
of the present universe ,
one has $l_H=\gamma/H_0$.
Making use of the continuity conditions of
$a(\tau)$ and of $a(\tau)'$ at the eight given joining points
$\tau_1$, $\tau_s$, $\tau_q$, $\tau_x$, $\tau_y$, $\tau_z$,
$\tau_2$, $\tau_E$, and $\tau_H$,
all parameters are fixed as the following:
\begin{eqnarray} \label{aling1}
&&\tau_a-\tau_E=\zeta_E^{\frac{1}{\gamma}},\nonumber\\
&&\tau_E -\tau_m=\frac{2}{\gamma}\,
                \zeta_E^{\frac{1}{\gamma}},\nonumber\\
&&\tau_2-\tau_m=\frac{2}{\gamma}\,
      \zeta_2^{-\frac{1}{2}}\zeta_E^{\frac{1}{\gamma}},\nonumber\\
&&\tau_2-\tau_g=\frac{1}{\gamma}\,\zeta_2^{-\frac{1}{2}}
      \zeta_E^{\frac{1}{\gamma}}, \nonumber\\
&&\tau_z-\tau_g=\frac{1}{\gamma}\,\zeta_z^{-1}\zeta_2^{-\frac{1}{2}}
      \zeta_E^{\frac{1}{\gamma}}, \nonumber\\
&&\tau_z-\tau_v=\frac{1}{\gamma}(1+v)\,\zeta_z^{-1}\zeta_2^{-\frac{1}{2}}
      \zeta_E^{\frac{1}{\gamma}}, \nonumber\\
&&\tau_y-\tau_v=\frac{1}{\gamma}(1+v)\,\zeta_y^{-\frac{1}{1+v}}\zeta_z^{-1}
      \zeta_2^{-\frac{1}{2}}\zeta_E^{\frac{1}{\gamma}}, \nonumber\\
&&\tau_y-\tau_f=\frac{1}{\gamma}\,\zeta_y^{-\frac{1}{1+v}}\zeta_z^{-1}
      \zeta_2^{-\frac{1}{2}}\zeta_E^{\frac{1}{\gamma}}, \nonumber\\
&&\tau_x-\tau_f=\frac{1}{\gamma}\,\zeta_x^{-1}
      \zeta_y^{-\frac{1}{1+v}}\zeta_z^{-1}
      \zeta_2^{-\frac{1}{2}}\zeta_E^{\frac{1}{\gamma}},\nonumber\\
&&\tau_x-\tau_n=\frac{1}{\gamma}(1+n)\,
      \zeta_x^{-1}
      \zeta_y^{-\frac{1}{1+v}}\zeta_z^{-1}
      \zeta_2^{-\frac{1}{2}}
      \zeta_E^{\frac{1}{\gamma}},\nonumber\\
&&\tau_q-\tau_n=\frac{1}{\gamma}(1+n)\,
      \zeta_q^{-\frac{1}{1+n}}\zeta_x^{-1}
      \zeta_y^{-\frac{1}{1+v}}\zeta_z^{-1}
      \zeta_2^{-\frac{1}{2}}\zeta_E^{\frac{1}{\gamma}}, \nonumber\\
&&\tau_q-\tau_e=\frac{1}{\gamma}\,
      \zeta_q^{-\frac{1}{1+n}}\zeta_x^{-1}
      \zeta_y^{-\frac{1}{1+v}}\zeta_z^{-1}
      \zeta_2^{-\frac{1}{2}}\zeta_E^{\frac{1}{\gamma}},\nonumber\\
&&\tau_s-\tau_e=\frac{1}{\gamma}\,
      \zeta_s^{-1}\zeta_q^{-\frac{1}{1+n}}\zeta_x^{-1}
      \zeta_y^{-\frac{1}{1+v}}\zeta_z^{-1}
      \zeta_2^{-\frac{1}{2}}\zeta_E^{\frac{1}{\gamma}},\nonumber\\
&&\tau_s-\tau_p=\frac{1}{\gamma}|1+\beta_{s}|\,
      \zeta_s^{-1}\zeta_q^{-\frac{1}{1+n}}\zeta_x^{-1}
      \zeta_y^{-\frac{1}{1+v}}\zeta_z^{-1}
      \zeta_2^{-\frac{1}{2}}\zeta_E^{\frac{1}{\gamma}},\nonumber\\
&&\tau_1-\tau_p=\frac{1}{\gamma}|1+\beta_{s}|\,
      \zeta_1^{-\frac{1}{1+\beta_{s}}}
      \zeta_s^{-1}\zeta_q^{-\frac{1}{1+n}}\zeta_x^{-1}
      \zeta_y^{-\frac{1}{1+v}}\zeta_z^{-1}
      \zeta_2^{-\frac{1}{2}}\zeta_E^{\frac{1}{\gamma}},\nonumber\\
&&\tau_1=\frac{1}{\gamma}(1+\beta)\,
      \zeta_1^{-\frac{1}{1+\beta_{s}}}\zeta_s^{-1}
      \zeta_q^{-\frac{1}{1+n}}\zeta_x^{-1}
      \zeta_y^{-\frac{1}{1+v}}\zeta_z^{-1}
      \zeta_2^{-\frac{1}{2}}\zeta_E^{\frac{1}{\gamma}},\nonumber\\
\end{eqnarray}
and
\begin{eqnarray} \label{aling2}
&&a_m=l_H\,\frac{\gamma^2}{4}\,\zeta_E^{-(1+\frac{2}{\gamma})},\nonumber\\
&&a_g=l_H\,\gamma\,\zeta_2^{-\frac{1}{2}}
    \zeta_E^{-(1+\frac{1}{\gamma})},\nonumber\\
&&a_v=l_H\,\gamma^{1+v}|1+v|^{-(1+v)}\,
    \zeta_z^{v}\zeta_2^{\frac{v-1}{2}}
    \zeta_E^{-(1+\frac{1+v}{\gamma})},\nonumber\\
&&a_f=l_H\,\gamma\frac{10}{11}\,\zeta_2^{-\frac{1}{2}}
    \zeta_E^{-(1+\frac{1}{\gamma})},\nonumber\\
&&a_n=l_H\,\frac{10}{11}\,\gamma^{1+n}|1+n|^{-(1+n)}
    \zeta_x^{n}\zeta_y^{\frac{n}{1+v}}
    \zeta_z^{n}\zeta_2^{\frac{n-1}{2}}
    \zeta_E^{-(1+\frac{1+n}{\gamma})},\nonumber\\
&&a_e=l_H\,\gamma\frac{25}{33}\,\zeta_2^{-\frac{1}{2}}
    \zeta_E^{-(1+\frac{1}{\gamma})},\nonumber\\
&&a_z=l_H\,\frac{25}{33}\,\gamma^{1+\beta_s}|1+\beta_s|^{-(1+\beta_s)}
    \zeta_s^{\beta_s}\zeta_q^{\frac{\beta_s}{1+n}}\zeta_x^{\beta_s}
    \zeta_y^{\frac{\beta_s}{1+v}}\zeta_z^{\beta_s}
    \zeta_2^{\frac{\beta_s-1}{2}}
    \zeta_E^{-(1+\frac{1+\beta_s}{\gamma})},\nonumber\\
&&l_0=l_H\,\frac{25}{33}\,\gamma^{1+\beta}\,|1+\beta|^{-(1+\beta)}
    \zeta_1^{\frac{\beta-\beta_s}{1+\beta_s}}
    \zeta_s^{\beta}\zeta_q^{\frac{\beta}{1+n}}\zeta_x^{\beta}
    \zeta_y^{\frac{\beta}{1+v}}\zeta_z^{\beta}
    \zeta_2^{\frac{\beta-1}{2}}
    \zeta_E^{-(1+\frac{1+\beta}{\gamma})}.
\end{eqnarray}

In the expanding universe,
the physical wavelength is related to the
comoving wavenumber $k$ by
\be
\lambda\equiv \frac{2\pi a(\tau)}{k},
\ee
 and the wavenumber $k_H$
corresponding to the present Hubble
radius is
\be
k_H = \frac{2\pi a(\tau_H)}{1/H_0}=2\pi \gamma.
\ee
There is another wavenumber involved
\be \label{ke}
k_E \equiv \frac{2\pi
a(\tau_E)}{1/H_0}=\frac{k_H}{1+z_E},
\ee
whose corresponding
wavelength at the time $\tau_E$ is the Hubble radius $1/H_0$.
In the
present universe the physical frequency corresponding to a
wavenumber $k$ is given by
\be \label{12}
\nu = \frac{1}{\lambda}=
\frac{k}{2\pi a (\tau_H)} = \frac{H_0}{2\pi\gamma} k.
\ee

\

\begin{center}
{\bf 3. Analytical solution of RGWs}
\end{center}

In the presence of the gravitational waves,
the perturbed metric  is
\be ds^2=a^2(\tau)[-d\tau^2+(\delta_{ij}+h_{ij})dx^idx^j],
\ee
where
the tensorial perturbation $h_{ij}$ is a $3\times 3$ matrix and is
taken to be transverse and traceless
\begin{eqnarray}
h^i_{\,\,i}=0,&&h_{ij,j}=0.
\end{eqnarray}
The wave equation of RGWs is
\be \label{eq1}
 \partial_{\nu}( \sqrt{-g}\partial^{\nu} h_{ij} )=0.
\ee
We decompose $h_{ij}$  into the Fourier
modes of the comoving wave number $k$ and into the polarization
state $\sigma$ as
\be \label{planwave}
h_{ij}(\tau,{\bf x})=
\sum_{\sigma}\int\frac{d^3k}{(2\pi)^3}
\epsilon^{\sigma}_{ij}h_k^{(\sigma)}(\tau) e^{i\bf{k}\cdot{x}} \, ,
\ee
where $h_{-k}^{(\sigma)*}(\tau)=h_k^{(\sigma)}(\tau)$ ensuring
that $h_{ij}$ be real,
$\epsilon^{\sigma}_{ij}$ is the polarization
tensor,
and $\sigma$ denotes the polarization states $\times,+$.
Here $h_{ij}$ is  treated as a classical field.
In terms of the mode
$h^{(\sigma)}_{k}$, Eq.(\ref{eq1}) reduces to
\be \label{eq}
h^{ (\sigma) }_{k}{''}(\tau) +2\frac{a'(\tau)}{a(\tau)}h^{ (\sigma)
}_k {'}(\tau) +k^2h^{(\sigma)}_k(\tau)=0.
\ee
By assumption,
 for each polarization, $\times$,  $+$,
the wave equation is the same and has
the same statistical properties,
so that the super index $(\sigma)$
can be dropped from $h^{(\sigma)}_k$ from now on.
Since
for all the stages of
expansion the  scale factor is of a power-law form
\be
a(\tau) \propto \tau^\alpha
\ee
the solution to
Eq.(\ref{eq}) is a linear combination of Bessel function $J_{\nu}$
and Neumann function $N_{\nu}$ \be \label{hom}
h_k(\tau)=\tau^{\frac{1}{2}-\alpha}
 \Big[a_1 J_{\alpha-\frac{1}{2}}(k \tau)
+a_2   N_{\alpha-\frac{1}{2}}(k \tau)\Big], \ee where the constants
$a_1$ and $a_2$ are completely determined by the continuity of $h_k$
and of $h'_k$ at the joining points
$\tau_1$, $\tau_s$, $\tau_q$, $\tau_x$, $\tau_y$, $\tau_z$,
$\tau_2$, and $\tau_E$.

In particular, we write down explicitly the solution for
the inflationary stage
since it give the initial condition for the spectrum of RGWs,
\be \label{infl}
h_k(\tau)=A_0 l_0^{-1}|\tau|^{-\frac{1}{2}-\beta} \Big[ A_1
J_{\frac{1}{2}+\beta}(x) +A_2 J_{-(\frac{1}{2}+\beta)}(x) \Big],
\,\,\,\, -\infty<\tau\leq \tau_1
\ee
where
$x\equiv k\tau$ and
\be
A_1=-\frac{i}{\cos \beta\pi}\sqrt{\frac{\pi}{2}}e^{i\pi\beta/2},
\,\,\,\,\,
A_2=\frac{1}{\cos\beta\pi}\sqrt{\frac{\pi}{2}}e^{-i\pi\beta/2},
\ee
are taken
\cite{grishchuk1993},
so that the so-called \textit{adiabatic
vacuum} is achieved: $\lim_{k\rightarrow \infty}h_k(\tau)\propto
e^{-ik\tau} $ in the high frequency limit \cite{parker}.
Moreover,
the constant $A_0$ in Eq.(\ref{infl}) is independent of $k$, whose
value  is determined by the initial amplitude of the spectrum.
For $k\tau\ll 1$ the $k$-dependence of $h_k(\tau)$ is given by
\be
h_k(\tau)\propto J_{\frac{1}{2}+\beta}(x)\propto
 k^{\frac{1}{2}+\beta}.
\ee
As will be seen later,
this choice will lead to the required
scale-invariant initial spectrum in Eq.(\ref{initialspectrum}).

It should be mentioned  that around the temperature $T \sim 2$ Mev
neutrinos decoupled from electrons and photons
and started free-streaming in space.
This will give rise to an anisotropic part  of the
energy-momentum tensor as a source of the equation of RGWs.
The previous works have shown that the neutrino free-streaming
in combination with the dark energy
would cause a reduction of the  amplitude of RGWs by $\sim 20\%$
in the low frequency range $(10^{-16}\sim 10^{-10})$ Hz
\cite{weinberg,Miao,yuki}.
Although this modified RGWs will contribute to
the CMB anisotropies and polarizations on very large scales,
the frequency range  is outside the frequency bands of LIGO and LISA.
On the other hand,
as will be seen,  the impact on RGWs
by the QCD transition is in the high frequency range  $\nu > 10^{-9}$ Hz.
Since these two frequency ranges are  not overlapped,
for simplicity of computing,
we will not include the neutrino free-streaming  here.

\begin{figure}
\centerline{\includegraphics[width=8cm]{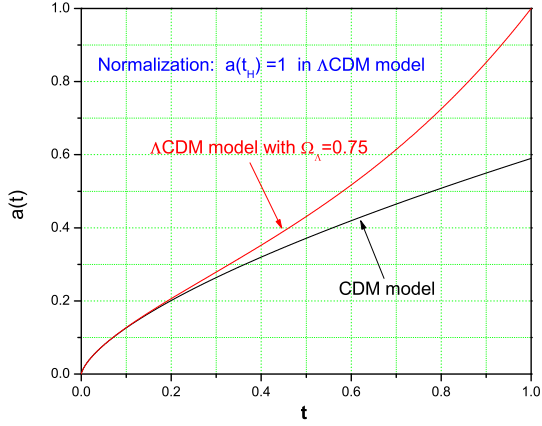}}
\caption{\label{fig2}
The scale factor $a(t)$
in the accelerating, and non-accelerating models, respectively.
Note that the horizontal axis is the cosmic time $t$.
At the present time $t=1$  the ratio of scale factors is
$a_{\Lambda CDM}/a_{CDM}\simeq 1.8$.}
\end{figure}

In our previous study  \cite{Miao},
we have examined the issue of how the RGWs would  be,
if our current universe were  matter-dominated.
The amplitudes of RGWs in the $\Lambda$CDM
(accelerating) and in the CDM  universe were compared,
and the ratio was found to be
$h_k(\tau_H)_{\Lambda CDM}/h_k(\tau_H)_{CDM} \sim 1.3$.
Here a similar examination is extended to
the case including the QCD transition and the $e^+e^-$ annihilation.
To be specific,
we assume that both universes  have the
same initial $a(\tau_2)$ and $a'(\tau_2)$
at the time $\tau_2$ with $z\simeq 3454$,
when $\rho_\Lambda \ll \rho_m=\rho_r$.
By numerically
solving the Friedmann equation in both models,
we plot the scale factor $a(t)$
in Fig.\ref{fig2},
showing that the ratio of scale factors at present
is $a(\tau_H)_{\Lambda CDM}/a(\tau_H)_{CDM}\simeq 1.8$.
As is known \cite{grishchuk, zhang2},
for wavelengths shorter
than the horizon  the modes decay as $h_k(\tau)\propto 1/a(\tau)$,
so the CDM model
would predict an amplitude of RGWs
higher than the $\Lambda$CDM model.
This is indeed confirmed by our calculation
including  the QCD transition and the $e^+e^-$ annihilation,
and the ratio is
\be \label{ratio}
h_k(\tau_H)_{CDM}/h_k(\tau_H)_{\Lambda CDM} \sim 1.8.
\ee
Moreover,
there are some subtleties with the matter-dominant model,
regarding to interpreting the current
observations.
As it stands, the actual universe is $\Lambda$CDM,
so the observed Hubble constant is properly interpreted as the
current expansion rate in the accelerating model,
$H_0=(a'/a^2)_{\tau_H}$.
The virtual
matter-dominant universe would
have a smaller rate $(a'/a^2)_{\tau_H} \simeq 0.4\, H_0$.
If the observed Hubble constant $H_0$
were regarded as the expansion rate of
the virtual matter-dominant universe \cite{yuki},
one would come up with an amplitude of $h_k(\tau_H)$ lower
by an extra factor $\sim 1.8$ than it should be.

\

\begin{center}
 {\bf 4. Spectrum of relic gravitational waves}
\end{center}

The spectrum of RGWs $h(k,\tau)$ at a time $\tau$
is defined by the
following equation \cite{grishchuk2000}:
\be
\int_0^{\infty}h^2(k,\tau)\frac{dk}{k}\equiv\langle0|
h^{ij}(\textbf{x},\tau)h_{ij}(\textbf{x},\tau)|0\rangle,
\ee
where
the right-hand side is the expectation value of the $h^{ij}h_{ij}$.
Calculation  yields the spectrum at present
\be \label{33}
h(k,\tau_H)  =
\frac{2}{\pi}k^{3/2}|h_k(\tau_H)|,
\ee
where the factor $2$
counts for the two independent polarizations.

One of the most important properties of
the inflation is that the initial spectrum of RGWs at the time
$\tau_i$ of the horizon-crossing during the inflation is nearly
scale-invariant \cite{grishchuk2000}:
\be \label{initialspectrum}
h(k,\tau_i) =A\Big(\frac{k}{k_H}\Big)^{2+\beta},
\ee
where
$2+\beta\simeq 0$, and $A$ is a constant to be fixed
by the observed CMB anisotropies  in practice.
The First Year WMAP
gives the scalar spectral index $n_s=0.99\pm 0.04$,
the Three Year WMAP gives  $n_s=0.951^{+0.015}_{-0.019}$
\cite{Spergel},
while in combination with constraints from  SDSS,
SNIa, and the galaxy clustering,
it would give $n_s=0.965\pm 0.012$ (68\%  CL) \cite{seljak1}.
The five-year WMAP data give $n_s=0.963^{+0.014}_{-0.015} $ \cite{WMAP5},
and the WMAP data combined with Baryon Acoustic Oscillations and
Type Ia supernovae   give
$n_s =0.960^{+0.014}_{-0.013} $ ($95\%$ CL) \cite{WMAP5hinshaw}..
From  the relation $n_s=2\beta+5$ \cite{grishchuk, zhang2},
the inflation index $\beta=-2.02$ for $n_s=0.951$.
As mentioned earlier,
we will allow the parameter $\beta$ to take values $>-2$
to demonstrate the RGWs spectrum.
Note that the constant $A$ is directly proportional
to  $A_0$ in Eq.(\ref{infl}).
Since the observed CMB anisotropies \cite{Spergel} is $\Delta T/T
\simeq 0.37\times 10^{-5}$ at $l\sim 10$,
which corresponds to
anisotropies on scales of the Hubble radius $1/H_0$,
so, as in Refs.\cite{Miao},
we take the  normalization of the spectrum
\be \label{norml}
h(k_E,\tau_H)=0.37\times10^{-5}r^{\frac{1}{2}},
\ee
where
$k_E$ is defined in Eq.(\ref{ke}),
its corresponding  physical frequency being
$\nu_E = k_E/2\pi a(\tau_H)=H_0/(1+z_E)\sim 1 \times 10^{-18}$ Hz.
The tensor/scalar ratio $r$
can be related to the slow roll parameter $\epsilon$
in the scalar inflationary model
as $r=16\epsilon$  \cite{Liddle}.
However, the value of $r$ is model-dependent,
and  frequency-dependent \cite{zhao2,baskaran}.
This has long been known to be
a notoriously thorny issue \cite{seljak2}.
In our treatment, for simplicity,
$r$ is only taken as a
constant parameter for normalization of RGWs.
Currently,
only observational constraints on  $r$ have been given.
The Three Year WMAP constraint is $r<2.2 $ ($95\%$ CL) evaluated at
$k=0.002$ Mpc$^{-1}$, and the full WMAP constraint is $r<0.55 $
(95\% CL) \cite{page}.
The combination from such observations, as of
SDSS, 3-year WMAP,
supernovae SN, and galaxy clustering, gives an upper limit $r<0.22$
($95\%$ CL) \cite{seljak1}.
The five-year WMAP gives a limit
$r< 0.43 $ ($95\%$) for power-law models \cite{WMAP5},
and the WMAP data combined with Baryon Acoustic Oscillations and
Type Ia supernovae   give
$r< 0.20 $ ($95\%$ CL) \cite{WMAP5hinshaw}.
For concreteness, we take $r=0.22$.

The spectral energy density $\Omega_g(k)$ of the RGWs is given by
\be \label{32}
\Omega_g(k)=\frac{\pi^2}{3}h^2(k,\tau_H)\Big(\frac{k}{k_H}\Big)^2,
\ee
which follows from the definition of
the total energy density of RGWs \cite{grishchuk2000}
\be  \label{gwe}
\Omega_{GW} \equiv \frac{\rho_g}{\rho_c} =
\int_{k_{low}}^{k_{upper}} \Omega_g(k)\frac{dk}{k},
\ee
where
$\rho_g=\frac{1}{32\pi G}h_{ij,0}h^{ij}_{,0}$ is the energy density
of RGWs, and $\rho_c=3H_0^2/8\pi G$ is the critical energy density.
The integration in Eq.(\ref{gwe}) has the lower and upper limits,
$k_{low}$ and  $k_{upper}$, as the cutoffs of the wavenumber.
The corresponding  frequencies are
$\nu_{low} \simeq 2\times 10^{-18} {\, \rm Hz}$ and
$\nu_{upper} \simeq 10^{10} {\, \rm Hz}$.
Detailed analyses of these limits are given in Ref.\cite{Miao}.
In past,
in the absence of direct detection of RGWs,
the constraint on RGWs through
the energy density $\Omega_{GW}$
has been commonly used;
especially, a bound from the Big Bang  nucleosythesis (BBN)
\cite{maggiore}
\be \label{BBN}
\Omega_{GW}  h_0^2 <8.9\times 10^{-6}
\ee
has been frequently employed  in practice,
where $h_0\sim 0.7$ being the Hubble parameter \cite{Spergel}.

In the following we  give the resulting spectra
$h(k,\tau_H)$ and $\Omega_g(k)$ of RGWs,
demonstrate  their explicit dependence
upon the model parameters $\beta$, $\beta_s$, $ \gamma$, and the
modifications by the QCD transition and $e^+e^-$ annihilation.
Also we will compare it with the sensitivity of detections,
such as  LIGO and LISA.

\begin{figure}
\centerline{\includegraphics[width=8cm]{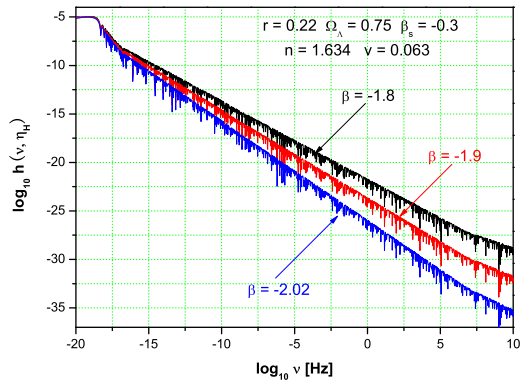}}
\caption{\label{fig3}
The spectrum $h(\nu, \tau_H)$ of GRWs is
very sensitive to the index $\beta$ of inflation. }
\end{figure}

Figure \ref{fig3} gives the spectrum $h(\nu, \tau_H)$ for
three values of the inflationary  index
$\beta = -1.8$, $-1.9$, and $-2.02$, respectively,
where the fixed $r=0.22$, $\Omega_\Lambda=0.75$,
$\beta_s=-0.3$, $n=1.634$, and $v=0.063$ are taken.
It is seen that $h(\nu,
\tau_H)$ is very sensitive to $\beta$.
A smaller $\beta$ will
generate lower amplitude of RGWs for all frequencies.

\begin{figure}
\centerline{\includegraphics[width=8cm]{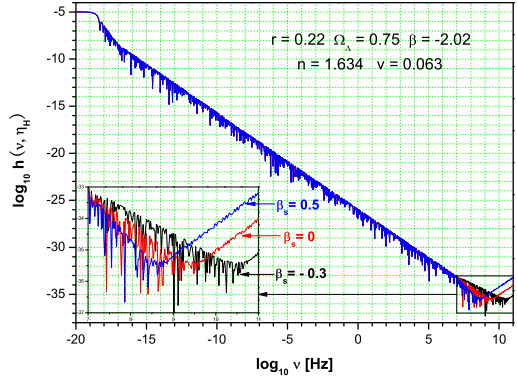}}
\caption{\label{fig4}
The reheating affects the spectrum
$h(\nu, \tau_H)$ only
in very high frequency range $\nu> 10^7$ Hz. }
\end{figure}

Figure \ref{fig4} shows
the influence of reheating stage on RGWs.
The spectrum is given for three different values of
$\beta_s=0.5$, $0$, and $-0.3$.
It is  seen  that the reheating process will affect RGWs only in
very high frequency range  $\nu > 10^7$ Hz.
This covers the frequency range of
some very high frequency gravity wave detection systems,
such as the Gaussian laser beam detector
aiming at the frequency $\nu\sim  10^{10}$ Hz ~\cite{fangyu},
or the  circular waveguide detector
aiming at $\nu> 10^{5}$ Hz \cite{cruise}.
We remark that the portion of predicted spectrum for
the frequency range $\nu> 10^{10}$ Hz may be not  reliable.
This is because
the energy scale of the conventional inflationary models
are usually less than $10 ^{16}$ Gev \cite{Miao},
which will gives a cutoff of frequency around
$\nu\sim 10^{10}$ Hz.

\begin{figure}
\centerline{\includegraphics[width=8cm]{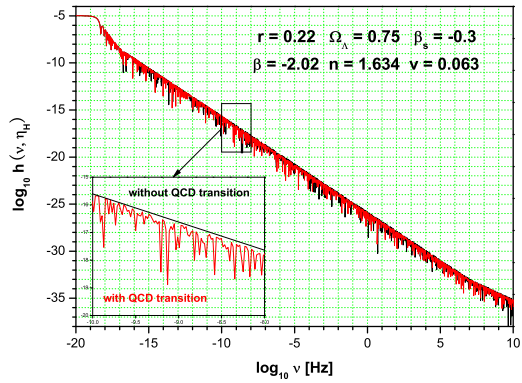}}
\caption{\label{fig5}
The QCD transition reduces $h(\nu, \tau_H)$ by
$\sim 20\%$ in the high frequency range $\nu > 10^{-9}$ Hz,
which covers the frequency band of LIGO and LISA.}
\end{figure}

Presented in Fig.\ref{fig5} is the modification of the spectrum
$h(\nu,\tau_H)$ by the QCD transition.
There is a critical frequency $\nu_Q\simeq 10^{-9}$ Hz,
below which the spectrum is the same for both models
with or without the QCD transition.
But, in the high frequency range $\nu> \nu_Q$ Hz,
the amplitude
of $h(\nu,\tau_H)$ is reduced by  $\sim 20\%$ in comparison
with the model without QCD transition.
\be
\label{1.2}
h_k(\tau_H)_{no\,\, QCD}/h_k(\tau_H)_{QCD} \simeq 1.2
\,\,\,\, {\rm for} \,\,\,\,  \nu> \nu_Q.
\ee
Note that
the frequency range of this reduction on RGWs
covers  those of the major laser interferometers GW detectors,
such as
LIGO and VIRGO operating effectively in the frequency bands
around $\sim 10^{2}$ Hz, and LISA around $\sim 10^{-3}$ Hz, respectively.
So the modifications on RGWs due to the QCD transition are relevant
to these  major detections.
In Appendix B we will give an interpretation
to the origin of this critical frequency $\nu_Q=10^{-9}$ Hz
and of the reduction in amplitude by $\sim 20 \%$
for $\nu>\nu_q$ Hz.

\begin{figure}
\centerline{\includegraphics[width=8cm]{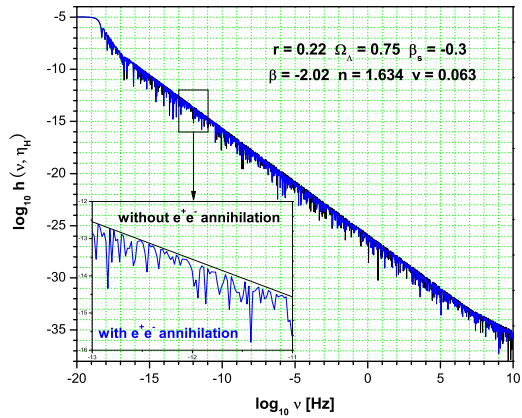}}
\caption{\label{fig6}
The $e^{+}e^{-}$ annihilation reduces $h(\nu, \tau_H)$ by
$\sim 10\%$ in the frequency range $\nu > 10^{-12}$ Hz,
which covers the frequency band of LIGO and LISA.}
\end{figure}

 Fig.\ref{fig6} gives the spectrum $h(\nu,\tau_H)$ modified
by the $e^{+}e^{-}$ annihilation,
which has a similar reduction effect.
The  lower frequency portion $\nu <  10^{-12}$ Hz
of the spectrum is not affected
by the $e^{+}e^{-}$ annihilation.
However, in the high frequency range $\nu>  10^{-12}$ Hz,
the amplitude of $h(\nu,\tau_H)$ is reduced by  $\sim 10\%$
by the  $e^{+}e^{-}$ annihilation.
Also the frequency range of this reduction
covers that of LIGO, VIRGO, LISA, etc.

\begin{figure}
\centerline{\includegraphics[width=8cm]{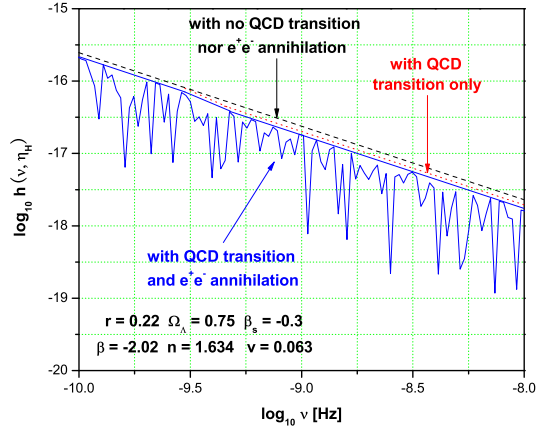}}
\caption{\label{fig7}
The spectrum $h(\nu,\tau_H)$ around $\nu = 10^{-9}$ Hz
for three cases: (1) with no QCD transition nor
$e^{+}e^{-}$ annihilation,
(2) with QCD transition only,
(3) with both QCD transition and $e^{+}e^{-}$ annihilation.}
\end{figure}

Fig.\ref{fig7} is an enlarged portion of the spectrum $h(\nu,\eta_H)$
around $\nu = 10^{-9}$ Hz,
illustrating the details of reductions of the spectrum
by both QCD transition and $e^{+}e^{-}$ annihilation.
The combination of the  QCD transition and the $e^{+}e^{-}$ annihilation
reduce $h(\nu,\tau_H)$ $\sim 30\%$.

\begin{figure}
\centerline{\includegraphics[width=8cm]{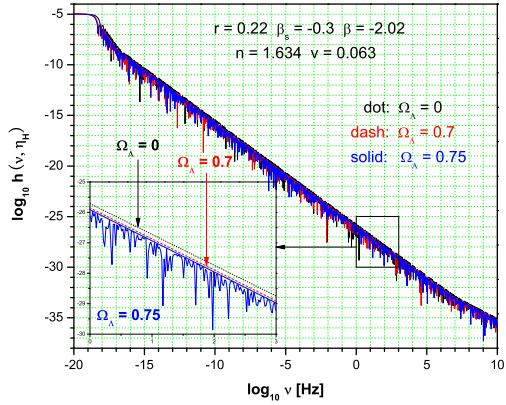}}
\caption{\label{fig8}
The spectrum  $h(\nu, \tau_H)$ depends upon
the dark energy $\Omega_\Lambda$ in the accelerating universe.
A larger $\Omega_\Lambda$ yields a lower $h(\nu,\tau_H)$.}
\end{figure}

The influence of the dark energy on
the spectrum $h(\nu,\tau_H)$ is demonstrated in Fig.\ref{fig8},
where $\Omega_{\Lambda}=0.0$,
$0.7$, and $0.75$ are taken respectively.
As explained in the last section,
over the whole range of frequency $(10^{-19} \sim 10^{10})$ Hz,
the amplitude of spectrum is suppressed by
the presence of $\Omega_\Lambda$, but the slope remains the same.
In particular, $h(\nu,\tau_H)$ with $\Omega_{\Lambda}=0.75$ is
reduced by $\sim 45\%$ in comparison with the  model $\Omega_{\Lambda}=0$.
(See Eq.(\ref{ratio}))

\begin{figure}
\centerline{\includegraphics[width=8cm]{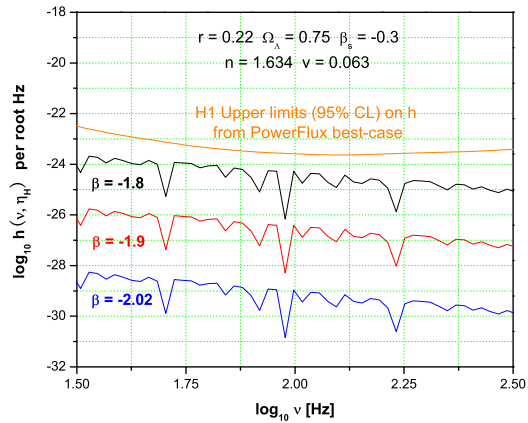}}
\caption{\label{fig9}
Comparison of the spectra with
the LIGO H1 Upper limits ($95\%$ CL) from PowerFlux
best-case in the Ref.\cite{Abbott}.
The vertical axis is the r.m.s
amplitude per root Hz defined in Eq.(\ref{rmsper}). }
\end{figure}

Figure \ref{fig9} is a comparison of the LIGO detection
with our calculated RGWs with the fixed tensor/scalar ratio $r=0.22$
and the dark energy $\Omega_\Lambda =0.75$.
The upper smooth curve is from the LIGO H1 Upper
limits ($95\%$ CL) from PowerFlux best-case \cite{Abbott},
and the lower three fluctuating curves are
the RGWs spectra of $\beta = -1.8$, $-1.9$, and $-2.02$,
corresponding to those in Fig.\ref{fig3}, respectively.
Here the vertical axis is the root
mean square amplitude per root Hz, which equals to
\be
\label{rmsper} \frac{h(\nu)}{\sqrt{\nu}}.
\ee
The plot gives only the frequency range  $(30 \sim  300)$ Hz,
on which the LIGO works efficiently.
The reductions due to the QCD transition and
the $e^{+}e^{-}$ annihilation have been incorporated in the curve of
calculated $h(\nu,\tau_H)/\sqrt{\nu}$.
Our result shows that
there is a gap of about one order of magnitude
even for the $\beta =-1.8  $ inflationary model.
As it currently stands,
the possibility for LIGO to detect the RGWs
predicted by the $\beta=-1.8$ inflationary model is not high,
let alone other models with $\beta<-1.8$.
Other two model parameters, i.e.,
the tensor/scalar ratio $r$  and the dark energy $\Omega_\Lambda$
will substantially influence the height of $h(\nu)$.
The five-year WMAP data improve the upper limit on
the tensor/scalar ratio $r<0.43$ ($95 \% $ CL)
for power-law inflationary models
and $r<0.58$ ($95 \% $ CL) for models with a running index,
and give the value of dark energy
$\Omega_\Lambda =0.721\pm 0.015$ \cite{WMAP5}.
So if we take the upper limit $r=0.43$ for power-law models,
the height of $h(\nu)$ will increase
by $\sim 40\%$ by Eq.(\ref{norml}),
and, furthermore, if we take $\Omega_\Lambda =0.721$,
$h(\nu)$ will increase by another $\sim 16\%$ \cite{zhang2}.
These together will allow an increase of $h(\nu)$
by a total $\sim 62\%$.
Therefore, the current LIGO with greatly enhanced
sensitivity \cite{ligo} will definitely be able
to put a constraint on the $\beta = -1.8$ inflationary model.
However,
it is seen from Fig.\ref{fig9} that
the curve for  $\beta \simeq -2.02 $,
supported by the scalar inflation models
and the WMAP data \cite{Spergel} \cite{WMAP5} \cite{WMAP5hinshaw},
is about five orders below the LIGO sensitivity.
So we may say that the RGWs generated
by scalar inflation models is unlikely to be directly detected by LIGO
at the moment.

\begin{figure}
\centerline{\includegraphics[width=8cm]{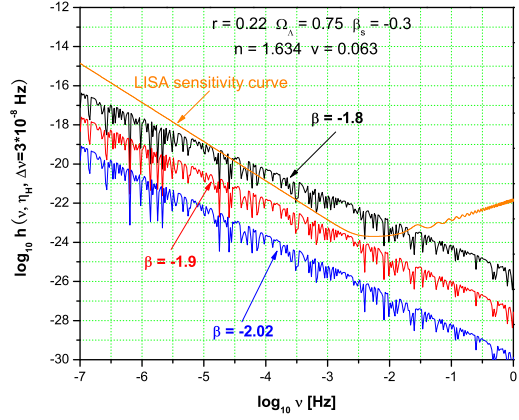}}
\caption{\label{fig10}
Comparison of the spectra with the
LISA sensitivity curve~\cite{shane}.
The vertical axis is the r.m.s
spectrum defined in Eq.(\ref{rmssp}).
The inflationary model of $\beta=-1.8$
can be tested by the LISA.}
\end{figure}

Figure~\ref{fig10} is a comparison of  the LISA
sensitivity curve with the spectra from  Fig.\ref{fig3}
in the lower frequency range $(10^{-7}, 10^0)$ Hz that
is also covered by the modifications of the QCD transition
and the $e^{+}e^{-}$ annihilation.
Assume that LISA has one year observation
time corresponding to frequency bin $\Delta\nu=3\times
10^{-18}$Hz (i.e., one cycle/year) around each frequency.
To
make a comparison with the sensitity curve,
we need to rescale the
spectrum $h(\nu)$ in Eq.(\ref{33}) into the root mean square
spectrum $h(\nu, \Delta\nu)$ in the band $\Delta \nu$
\cite{grishchuk2000} \cite{maggiore},
\be
\label{rmssp} h(\nu, \Delta\nu) =
h(\nu)\sqrt{\frac{\Delta\nu}{\nu}}.
\ee
This r.m.s spectrum can be
directly compared  with the 1 year integration sensitivity curve
that is downloaded from LISA \cite{shane}.
The plot shows that LISA
by its present design will be able to detect
the inflationary model of $\beta=-1.8$.
If the ratio $r > 0.22$, it is
still possible for LISA to detect the model of
$\beta= -1.9$.
However, LISA is unlikely to be able to directly
detect the model of $\beta=-2.02$, as there is a gap of two orders.
One may say that, in regards to detection of RGWs,
LISA will give a stronger constraint
on the RGWs spectrum than LIGO will do.

\begin{figure}
\centerline{\includegraphics[width=8cm]{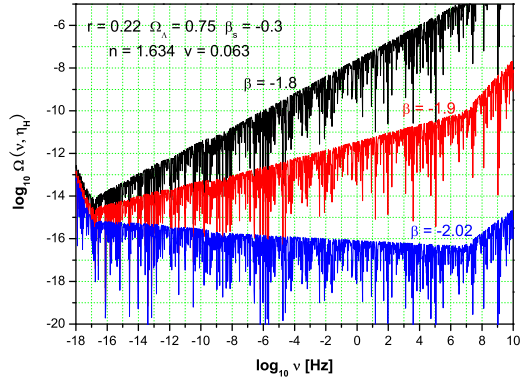}}
\caption{\label{fig11}
The spectral energy density
$\Omega_g(\nu)$ for various values of the parameter $\beta$.}
\end{figure}

Figure \ref{fig11} shows the $\beta$-dependence of
the spectral energy density $\Omega_g(\nu)$ defined in Eq.(\ref{32}).
Clearly, $\Omega_g$ is very sensitive to
the inflationary parameter $\beta$.
A larger $\beta$ gives a higher $\Omega_g$.
The Advanced LIGO~\cite{ligo} will be
able to detect RGWs with $\Omega_g h^2>10^{-9}$ at $\nu \sim 100$Hz,
and it might impose stronger constraints on other inflationary
models.
On the other hand,
the total energy density  $\Omega_{GW}$ defined in Eq.(\ref{gwe})
has also been used as a constraint on RGWs.
Taking the parameters $r=0.22$, $\Omega_\Lambda =0.75$,
$\beta_s=-0.3$, and $n=1.634$,
we find $\Omega_{GW}=1.11\times10^{-2}$ for the
inflationary model of $\beta=-1.8$,
which is $3$ orders higher than the BBN bound  \cite{maggiore}
in Eq.(\ref{BBN}).
So this model is disfavored,
 unless some other mechanism is introduced to reduce its $\Omega_{GW}$.
We also obtain $\Omega_{GW}=1.97\times10^{-8}$
for the model $\beta=-1.9$,
and $\Omega_{GW}= 1.41\times10^{-14}$ for the model $\beta=-2.02$;
both models are  safely below the BBN bound in Eq.(\ref{BBN}).

\begin{figure}
\centerline{\includegraphics[width=8cm]{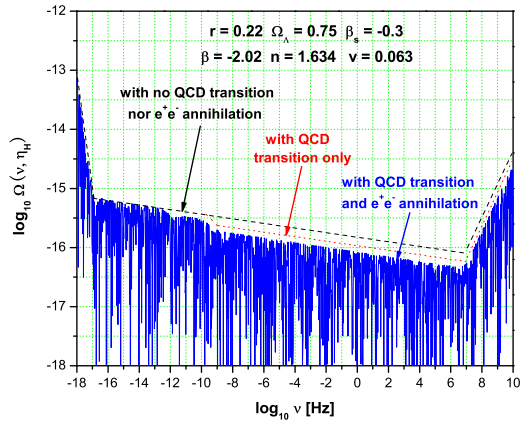}}
\caption{\label{fig12}
The spectral energy density $\Omega_g(\nu)$
for three cases: (1) with no QCD transition nor
$e^{+}e^{-}$ annihilation,
(2)  with QCD transition only,
(3) with both QCD transition and $e^{+}e^{-}$ annihilation.}
\end{figure}

Fig.\ref{fig12} shows  $\Omega_g(\nu)$ in the $\beta=-2.02$
modified by the QCD transition and $e^+e^-$ annihilation,
where the reduction on RGWs is more noticeable.
In the range $\nu> 10^{-9}$ Hz
the QCD transition alone reduces $\Omega_g(\nu)$ by $\sim  30\%$,
and the combination of QCD transition and the $e^{+}e^{-}$ annihilation
reduces $\Omega_g(\nu)$ by  $\sim 50\%$.

\begin{figure}
\centerline{\includegraphics[width=8cm]{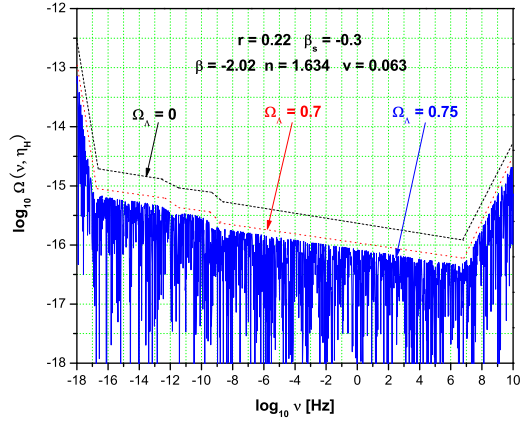}}
\caption{\label{fig13}
The spectral energy density $\Omega_g(\nu)$
for different $\Omega_\Lambda$.
A larger $\Omega_\Lambda$ yields a lower $\Omega_g(\nu)$.}
\end{figure}

The impact of dark energy on $\Omega_g(\nu)$
for the model $\beta=-2.02$ is plotted in Fig.\ref{fig13}.
Three values of $\Omega_\Lambda =0.0$, $0.70$, $0.75$ are taken.
A larger
$\Omega_\Lambda$ gives a lower $\Omega_g$.
It is seen that the $\Omega_\Lambda$
causes a decrease of the amplitude of $\Omega_g$ by
a factor of $1.6$ over the whole range of frequencies.

\

ACKNOWLEDGMENT:
We are grateful the referees for valuable suggestions.
Y.Zhang's research work is supported by the CNSF
No.10773009, SRFDP, and CAS.

\begin{center}
{\bf Appendix A: Modelling of QCD transition and
$e^{+}e^{-}$ annihilation}
\end{center}

The detail of the QCD transition is notoriously complicated
and still under study.
Here we will only
consider the change of the effective degree of freedom,
and give a simple working model for the scale factor $a(\tau)$
around the QCD transition lasting a short period.
Thus the radiation era can be tentatively divided into
the three parts with the scale factor $a(t)$
been listed in Eqs.(\ref{rad1}), (\ref{rad2}), and (\ref{rad3}).
After using the continuity conditions
of $a(\tau)$ and of $a(\tau)'$ at the two given joining points
$\tau_x$ and $\tau_2$,
one still needs to determine $3$ parameters,
$\tau_q$, $\tau_x$, $n$.
Since the QCD transition temperature $T\propto 1/a$
during the radiation era, as soon as $T$ is given,
the initial time $\tau_q$ of QCD transition is determined,
so is the parameter $\zeta_s$.
Still  $n$ and $\tau_x$ need to be fixed in the following.

Around the QCD transition,
the contributions from the matter $\rho_m$
and the dark energy $\rho_{\Lambda}$
can be neglected.
Only the radiation is important,
of which the energy density, the pressure, and the entropy density
 are given by
\begin{equation} \label{A4}
\rho_r(T)=\frac{2\pi^{2}}{30}g_{\ast}(T)T^{4},
\end{equation}
\begin{equation} \label{A5}
p_r(T)=\frac{1}{3} \rho_r(T),
\end{equation}
\begin{equation} \label{A3}
s(T)=(\rho_r+p_r)/T=\frac{2\pi^{2}}{45}g_{\ast s}(T)T^{3},
\end{equation}
respectively,
where $g_{\ast}$ and $g_{\ast s}$ denote the effective number of
relativistic species contributing to the energy density
and entropy, respectively \cite{Kolb}.
In an adiabatic universe,
the entropy per unit comoving volume is conserved
\begin{equation} \label{A2}
S(T)=s(T)a^{3}(T)=constant.
\end{equation}
From Eqs. (\ref{A4}), (\ref{A3}) and  (\ref{A2}),
one has
\be  \label{A6}
\rho_r \propto g_{\ast}  g_{\ast s}^{-4/3}a^{-4}.
\ee
During the QCD transition era,
one has $g_{\ast}=g_{\ast s}$  \cite{schwarz,yuki}.
Therefore, here we will not distinguish
the difference between $g_{\ast}$ and $g_{\ast s}$.
Thus, Eq.(\ref{A6}) is reduced to
\be  \label{A7}
\rho_r \propto g_{\ast}^{-1/3}a^{-4},
\ee
which is plugged into the Friedmann equation
$ (\frac{ a'}{a^2})^{2}=\frac{8\pi G}{3}\rho_r$, yielding
\be \label{p}
(a') ^2 \propto g_*^{-1/3}.
\ee
Thus, with the decreasing of $g_*$ across the QCD transition
the expansion rate $a'$ in terms of the conformal time increases.
As predicted by the Standard Model of particle physics,
before the QCD transition,
$g_{\ast }(\tau_q)=51.25$ at the time $\tau_q$;
and after the QCD transition,
it becomes $g_{\ast}(\tau_x)=17.25$
at the time  $\tau_x$ \cite{Kampfer,schwarz2}.
Applying Eq.({\ref{p}}) to Eqs.(\ref{rad1}) and (\ref{rad3})
yields the ratio
\be \label{A8}
\frac{a_{f}}{a_{e}}=
\Big[\frac{ g_{\ast }(\tau_q) }{ g_{\ast}(\tau_x) }\Big]^{\frac{1}{6}}
               \simeq 1.2.
\ee
From the expression $\zeta_q=\frac{a(\tau_x)}{a(\tau_q)}$,
one has
\be
\label{kk}
\zeta_q^{\frac{n}{1+n}}=1.2.
\ee
If $\tau_x$ is determined (equivalent to giving $\zeta_q$),
Eq.(\ref{kk}) will fix the parameter $n$.
Let us determine $\zeta_q$.
The form of power-law in Eq.(\ref{rad2}) implies
$g_{\ast}\propto \tau^{-6n}$ during the QCD transition era \cite{yuki},
thus Eq.(\ref{A7}) reduces to
\be
\label{den}
\rho \propto \tau^{2n}a^{-4}.
\ee
In terms of the cosmic time $t$ defined by
$dt = a d\tau$,
Eq.(\ref{rad2}) implies
\be   \label{aa}
a\propto t^{\frac{1+n}{2+n}}.
\ee
Therefore, one gets
\be   \label{at}
\zeta_q\equiv\frac{a(\tau_x)}{a(\tau_q)}
=\Big(\frac{t_{x}}{t_{q}}\Big)^{\frac{1+n}{2+n}}.
\ee
Once
the ratio of $t_{x}/t_{q}$ is given,
then two parameters $n$ and $\zeta_q$ can be fixed from
Eqs.(\ref{kk}) and (\ref{at}) immediately.
Ref.\cite{Kampfer} gives a value of the ratio  $t_{x}/t_{q}\sim  2$;
Ref.\cite{schwarz2} estimated
the typical duration of QCD transition
to be the order  of $ \sim 0.1 t_{q}$,
i.e., $t_{x}/t_{q} \sim  1.1$.
For concreteness in actual calculation,
we take
\be
t_{x}=(1.3\sim 2.0)t_{q}.
\ee
In Table \ref{QCD transition era}
the values of $n$ and  $\xi_q$ are listed
for each given value of $t_x/t_q$.
Here our treatment is different from that in Ref.\cite{yuki},
where the  choice  of  $n$ was put in by hand.

\begin{table}
\caption{The Values of $n$ and  $\xi_q$ for different
 $\frac{t_x}{t_q}$ (T=190MeV)}
\begin{center}
\label{QCD transition era}
\begin{tabular}{|c|c|c|c|}
  \hline
  $t_x/t_q$ & $n$ & $\zeta_{q}$ \\
  \hline
  1.3  & 4.556  & 1.25  \\
  1.5  & 1.634  & 1.34  \\
  1.8  & 0.899  & 1.47  \\
  2.0  & 0.714  & 1.55  \\
  \hline
\end{tabular}
\end{center}
\end{table}

In our model,
there are three physical parameters:
the QCD transition temperature $T$
(i.e. the beginning time $t_q$),
the duration of transition (i.e. the ratio $t_x/t_q$),
and the change of $g_*$
(i.e. the behavior of $a(\tau)$ during the transition).
As our calculation shows,
$\Omega_g(\nu)$
is not very sensitive to the concrete values of  $T$ and of $t_x/t_q$.
For  $T=160 $ and $190$ Mev,
 the difference in $\Omega_g(\nu)$
is only $\sim 1\%$,
and, similarly, for $t_x/t_q =1.3$ and $2$
the difference in $\Omega_g(\nu)$ is only $\sim 1\%$.
This small difference is also reflected by
the total energy density
$\Omega_{GW}$ for different $T$ and $t_x/t_q$
listed in Table \ref{QCD detail}
for $r=0.22$, $\Omega_\Lambda=0.75$, $\beta=-2.02$,
and $\beta_s=-0.3$.
Therefore, as a conclusion,
the most important element affecting the spectrum of RGWs
is the change of $g_*$,
which are completely determined
by the Standard Model of particle physics.

\begin{table}
\caption{$\Omega_{GW}$ for different QCD transition models}
\begin{center}
\label{QCD detail}
\begin{tabular}{|c|c|c|}
\hline
with QCD transition& $ \Omega_{GW}(160MeV)$ & $\Omega_{GW}( 190MeV)$ \\
\hline
$t_x/t_q=1.3$ & $1.40\times10^{-14}$ & $1.44\times10^{-14}$ \\
$t_x/t_q=1.5$ & $1.38\times10^{-14}$ & $1.41\times10^{-14}$ \\
$t_x/t_q=2.0$ & $1.35\times10^{-14}$ & $1.36\times10^{-14}$ \\
\hline \hline
without QCD trnsition & $1.54\times10^{-14}$ & $1.54\times10^{-14}$ \\
\hline
\end{tabular}
\end{center}
\end{table}

The same treatment can be applied to the $e^{+}e^{-}$ annihilation.
According to the Standard Model of particle physics,
before the $e^{+}e^{-}$ annihilation,
$g_{\ast }(\tau_y)=g_{\ast s}(\tau_y)=10.75$ at the time $\tau_y$;
and after the $e^{+}e^{-}$ annihilation,
$g_{\ast}(\tau_z)=3.36$ and $g_{\ast s}(\tau_z)=3.91$
at the time  $\tau_z$ \cite{schwarz,yuki}.
Applying Eq.({\ref{A6}}) to Eqs.(\ref{rad3}) and (\ref{rad5})
yields the ratio
\be \label{A9}
\frac{a_{g}}{a_{f}}=\frac{g_{\ast}^{-1/2}
             (\tau_y)\,\, g_{\ast s}^{2/3}(\tau_y)}
{g_{\ast}^{-1/2}(\tau_z)\,\, g_{\ast s}^{2/3}(\tau_z)}\simeq 1.1.
\ee
One can obtain
\be
\label{ll}
\zeta_y^{\frac{v}{1+v}}=1.1.
\ee
Since $T\propto 1/a$ and
the $e^{+}e^{-}$ annihilation occurred during
$T\simeq (0.5 \sim 0.1)$ MeV  \cite{yuki},
one obtains $\zeta_y\equiv\frac{a(\tau_z)}{a(\tau_y)}=5$
and $v=0.063$ that are used in Eq.(\ref{rad4}).

\begin{center}
{\bf Appendix B: Simple interpretation of modifications
 of RGWs by QCD transition and $e^+e^-$ annihilation}
\end{center}

\begin{figure}
\centerline{\includegraphics[width=8cm]{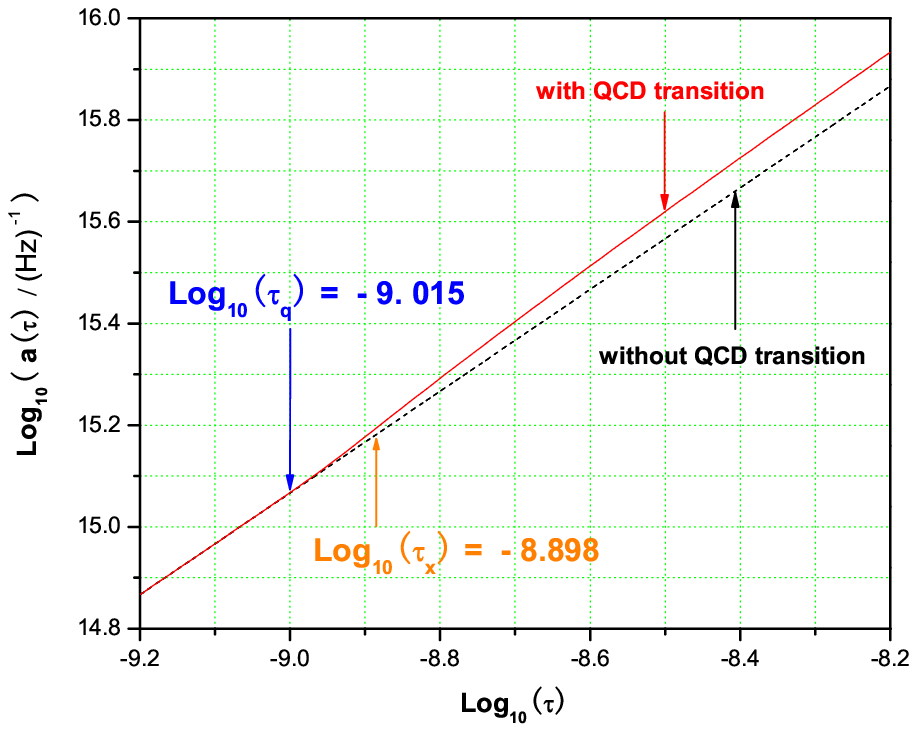}} \caption{
\label{fig14} The scale factor $a(\tau)$ is affected by the QCD
transition. For $\tau < 10^{-9}$, $a(\tau)_{QCD}=a(\tau)_{no \,
QCD}$. But for $\tau > 10^{-9}$, $a(\tau_{H})_{QCD}/a(\tau_{H})_{no
\, QCD}\simeq1.2$.}
\end{figure}

Given the wave equation in Eq.(\ref{eq}),
there are two limiting cases for wavelength.
For the short wavelength $k\gg \frac{a'}{a}$,
the mode function $h_k$ has a decreasing amplitude,
\be \label{mo1}
h_k (\tau)\propto 1/a(\tau).
\ee
and, for the long wavelength $k\ll \frac{a'}{a}$,
the mode $h_k$ is simply a constant
\begin{equation}
\label{mo2}
h_k (\tau) \simeq C_k.
\end{equation}
This property itself will be able to qualitatively  account for
why the QCD transition induces a decrease in
the spectrum $h(\nu, \tau_H)$ in Eq.(\ref{1.2}).
In Fig.\ref{fig14} the scale factor $a(\tau)$
is plotted for the QCD transition era around $T\sim 190$ Mev,
and it is seen that
\be
a(\tau)_{QCD}=a(\tau)_{no \,QCD}, \,\,\,\,\,\, \tau < 10^{-9},
\ee
\be \label{a1}
\frac{a(\tau)_{QCD}}{a(\tau)_{no \,QCD}}
\simeq1.2, \,\,\,\,\,\, \tau > 10^{-9}.
\ee
This result,
together with Eq.(\ref{mo1}), explains the reduction
of  $h(\nu, \tau_H)$ by  $\sim 20\%$ by the QCD transition,
as explicitly shown in Fig.\ref{fig5}.
We need to explain why
the reduction occurs for  $\nu> \nu_Q= 10^{-9} $ Hz.
Since at any time $\tau$ those  modes with
\be
k > \frac{a'(\tau)}{a(\tau)}\simeq \frac{1}{\tau}
\ee
will  decay as in Eq.(\ref{mo1}),
  one sees that only those modes with
\be \label{crit}
k>  k_Q \equiv \frac{1}{\tau_q}
\ee
will be reduced by the QCD  transition.
According to Eq.(\ref{12}),
in the present universe
the physical frequency corresponding to $k_Q$
is
\be \label{criticalfrequency}
\nu_Q = \frac{k_Q}{2\pi a(\tau_H)}
 = \frac{H_0}{2 \pi \gamma \tau_q}
 \simeq 10^{-9} \,\,{\rm Hz}
\ee
for $\gamma = 1.044$ and
the Hubble constant  $H_0 \simeq 2.36\times 10^{-18} $ Hz.
This  critical frequency $\nu_Q$
in Eq.(\ref{criticalfrequency})
matches that demonstrated in
Fig.\ref{fig5} and Fig.\ref{fig12},
and is about 2 orders lower than $10^{-7}$ Hz,
a value  quoted in Ref.~\cite{schwarz}.
Thus, the features in the calculated spectrum of RGWs
have been fully explained.
We remark that a higher temperature $T$ of the QCD transition
would yield an earlier time $\tau_q$
and a higher critical frequency $\nu_Q$.
As for the reduction by $e^+e^-$ annihilation,
the interpretation is similar to the above.



\small
\begin{thebibliography}{80}

\bibitem[*]{email}{e-mail: yzh@ustc.edu.cn}

\bibitem{grishchuk} L. P. Grishchuk,
     Sov.Phys.JETP \textbf{40}, 409  (1975);
       Class.Quant.Grav.\textbf{14}, 1445 (1997);
       astro-ph/0008481.

\bibitem{starobinsky} A. A. Starobinsky,
JEPT Lett. \textbf{30} 682 (1979); Sov.Astron.Lett.{\bf 11}, 133
(1985);
V. A. Rubakov, M.Sazhin, and A. Veryaskin, Phys.Lett.B
\textbf{115}, 189 (1982);
R. Fabbri \& M.D. Pollock, Phys.Lett.B{\bf125}, 445 (1983);
L. F. Abbott \& M.B. Wise, Nucl.Phys.B{\bf244}, 541 (1984);
L. F. Abbott \& D. D. Harari, Nucl.Phys.B\textbf{264}, 487(1986);
B. Allen, Phys.Rev.D\textbf{37}, 2078 (1988);
V. Sahni, Phys.Rev.D\textbf{42}, 453 (1990);
A. Riazuelo \& J. P. Uzan, Phys.Rev.D\textbf{62}, 083506 (2000);
H. Tashiro, et.al, Class.Quant.Grav.\textbf{21}, 1761 (2004);
A. B. Henriques, Class.Quant.Grav.\textbf{21}, 3057 (2004);
L. A. Boyle and P. J. Steinhardt, arXiv:astro-ph/0512014



\bibitem{zhang2} Y. Zhang,  et.al.,
   Class. Quant. Grav. {\bf 22}, 1383 (2005);
   Chin. Phys. Lett. {\bf 22},  1817 (2005);
    Class. Quant. Grav.{\bf 23}, 3783 (2006).


\bibitem{ligo} http://www.ligo.caltech.edu/;
        http://www.ligo.caltech.edu/advLIGO.

\bibitem{virgo} http://wwwcascina.virgo.infn.it/.

\bibitem{geo} http://www.geo600.uni-hannover.de/.

\bibitem{tamma} http://tamago.mtk.nao.ac.jp/.

\bibitem{aigo} http://www.anu.edu.au/Physics/ACIGA/

\bibitem{lisa} http://lisa.nasa.gov/; http://www.lisa.caltech.edu/

\bibitem{astrod}
W.T. Ni, S. Shiomi and A.C. Liao, Class.Quant.Grav. {\bf 21} S641 (2004).

\bibitem{bbo}
http://universe.nasa.gov/program/bbo.html; ~~V. Corbin and
N.J. Cornish, Class.Quant.Grav. {\bf 23} 2435 (2006).

\bibitem{decigo}
N. Seto, S. Kawamura and T. Nakamura,
        Phys.Rev.Lett. {\bf 87} 221103 (2001).

\bibitem{basko} M.M. Basko \& A.G. Polnarev,
MNRAS {\bf191}, 207  (1980);
A.G. Polnarev,  AZh, {\bf62}, 1041 (1985);
N.  Kaiser,  MNRAS {\bf202}, 1169 (1983);
J.R. Bond \& G. Efstathiou, ApJ {\bf 285}, L45 (1984);
                           MNRAS {\bf 226}, 655 (1987);
R. Crittenden, R.L. Davis, P.J. Steinhardt, ApJ {\bf 417}, L13 (1993);
R.G. Crittenden, D. Coulson, and  N.G. Turok,
        Phys.Rev.D.{\bf 52}, R5402 (1995);
D. Coulson, R.G. Crittenden, and N.G. Turok, ApJ {\bf 417}, L13 (1993);
D.D. Harari \&  M. Zaldarriaga, Phys.Lett.B{\bf 319}, 96 (1993) ;
M. Zaldarriaga \& D.D. Harari, Phys.Rev.D{\bf 52}, 3276  (1995);
R.A. Frewin, A.G. Polnarev, P. Coles, MNRAS, 266, L21 (1994).
B.G. Keating, P.T. Timbie, A. Polnarev, and J. Steinberger,
        ApJ, {\bf 495}, 580 (1998);
A. Kosowsky,  Ann.Phys. {\bf 246}, 49 (1996);
M. Zaldarriaga \&  U. Seljak, Phys.Rev.D{\bf 55}, 1830 (1997);
U. Seljak \& M. Zaldarriaga, Phys.Rev.Lett.{\bf 78}, 2054 (1997).
W. Hu \& M. White, Phys.Rev.D{\bf 56}, 596 (1997).

\bibitem{Kamionkowski} M. Kamionkowski, A. Kosowsky,
and A. Stebbins, Phys.Rev.D{\bf 55}, 7368 (1997)

\bibitem{pritchard} J. R. Pritchard \& M. Kamionkowski,
Annals. Phys.{\bf 318},  2  (2005).

\bibitem{zhao2} W. Zhao and Y. Zhang, Phys.Rev.D{\bf 74}, 083006 (2006).

\bibitem{baskaran} D. Baskaran, L.P. Grishchuk,
    and A.G. Polnarev, Phys. Rev. D{\bf 74}, 083008 (2006);
 Polnarev A.G., Miller N.J.,  Keating B.G.,
              arXiv: astro-phy 0710.3649.



\bibitem{Spergel} D.N. Spergel, et al,  ApJS {\bf 148}, 175 (2003).

         D.N. Spergel, et.al. ApJS {\bf 170}, 377 (2007).

\bibitem{Tegmark} M. Tegmark et.al. Phys.Rev. D{\bf 74}, 123507 (2006)

\bibitem{seljak1} U. Seljak, A. Slosar and P. McDonald,
                 JCAP {\bf 0610}, 014 (2006).

\bibitem{page} L. Page, et.al. ApJS {\bf 170}, 335 (2007).

\bibitem{hinshaw} G. Hinshaw et.al. ApJS {\bf 170}, 263 (2007).



\bibitem{zhao}  T. L. Smith, M. Kamionkowski and A. Cooray,
                          Phys.Rev. D{\bf73},  023504 (2006);
                W. Zhao and Y. Zhang, Phys.Rev.D{\bf 74}, 043503 (2006).

\bibitem{perlmutter} A.G. Riess,  {\it et al.},
               Astron.J. {\bf 116}, 1009  (1998);
               ApJ. {\bf 117}, 707 (1999);
S.  Perlmutter {\it et al.}, ApJ {\bf 517}, 565 (1999);
J.L. Tonry  {\it et al.}, ApJ {\bf 594}, 1 (2003);
R.A. Knop {\it et al.}, ApJ {\bf 598}, 102 (2003);
A.G. Riess  {\it et al.}, Astron.J. {\bf 607}, 665 (2004).

\bibitem{weinberg} S. Weinberg,  Phys.Rev.D{\bf 69}, 023503 (2004);
       D. A. Dicus and  W. W. Repko,
           Phys.Rev.D{\bf 72}, 088302 (2005).

\bibitem{Miao} H.X. Miao and Y. Zhang, Phys.Rev.D{\bf 75}, 104009 (2007).

\bibitem{cheng} M. Cheng et al, Phys.Rev.D{\bf 74}, 054507 (2006).

\bibitem{karsch} F. Karsch, arXiv:0711.0661.

\bibitem{schwarz} D. J. Schwarz,
   Mod.Phys.Lett.A{\bf 13}, 2771 (1998).

\bibitem{yuki} Y. Watanabe and E. Komatsu,
   Phys.Rev.D{\bf 73},  123515 (2006).



\bibitem{grishchuk2000} L. P. Grishchuk,
   in {\it Lecture Notes in Physics}, Vol. 562, p.167,
   Springer-Verlag, (2001),  (gr-qc/0002035).

\bibitem{Liddle} A.R. Liddle and D.H. Lyth,
          Phys.Lett. B{\bf 291}, 391 (2006).

\bibitem{Zhang} Y. Zhang, Gen.Rel.Grav.{\bf 34}, 2155  (2002);
                         Gen.Rel.Grav.{\bf 35}, 689 (2003);
                         Chin.Phys.Lett.{\bf 20}, 1899 (2003);
                         Chin.Phys.Lett.{\bf 21}, 1183 (2004).
W. Zhao and Y. Zhang, Phys.Lett. B{\bf 640}, 69 (2006);
                    Class.Quant.Grav.{\bf23},  3405 (2006).
Y. Zhang, T.Y. Xia, and W. Zhao,
                    Class.Quant.Grav.{\bf 24},  3309 (2007).
T.Y. Xia and Y. Zhang,
                     Phys.Lett. B{\bf 656}, 19 (2007).

\bibitem{grishchuk1993} L. P. Grishchuk,
    Phys.Rev.D{\bf 48}, 3513 (1993).

\bibitem{parker} L. Parker, Phys.Rev.183, 1057 (1969).


\bibitem{WMAP5} J. Dunkley, et al,
                arXiv:astro-ph 0803.0586.

\bibitem{WMAP5hinshaw} G. Hinshaw, et al, arXiv:astro-ph 0803.0732;
                          E. Komatsu,, et al, arXiv:astro-ph 0803.0547

\bibitem{seljak2} U. Seljak, et.al., Phys.Rev. D{\bf 71} 103515 (2005);
      A. Cooray, P.S. Corasaniti,
              T. Giannantonio and A. Melchiorri,
              Phys. Rev. D{\bf 72}, 023514 (2005);
 T.L. Smith, M. Kamionkowski and A. Cooray,
          Phys. Rev. D{\bf 73}, 023504 (2006);arXiv: astro-ph 0802.1530;
 A. Linde, V. Mukhanov and M. Sasaki,
           JCAP {\bf 0510}, 002 (2005);
 V. Mukhanov and A.Vikman, JCAP {\bf 0602}, 004 (2006).


\bibitem{maggiore} M. Maggiore, Phys.Rept.{\bf 331}, 283 (2000);
R.H. Cyburt, J. Ellis, B.D. Fields, and K.A. Olive,
           Phys. Rev. D{\bf 67},  103521 (2003);
R.H. Cyburt et.al. Astropart.Phys.{\bf 23},  313 (2005);
T.L. Smith, E. Pierpaoli, and M. Kamionkowski,
          Phys.Rev.Lett{\bf 97}, 021301 (2006).


\bibitem{fangyu} F.Y. Li,  M.X.  Tang, and D.P. Shi,
       Phys.Rev.D{\bf 67}, 104008  (2003);
       M.L. Tong, Y. Zhang, and F.Y. Li, under preparation.

\bibitem{cruise}  A.M. Cruise, Class.Quant.Grav. {\bf 17}, 2525 (2000) ;
        A.M. Cruise and R.M.J. Ingley,
             Class.Quant.Grav.{\bf 22},  S479 (2005);
             Class.Quant.Grav.{\bf 23}, 6185 (2006).
        M.L. Tong and Y. Zhang, arXiv:gr-qc 0711.4909,
        to appear in ChJAA (2008).

\bibitem{Abbott} B. Abbott, et al. 2007 arXiv:0708.3818

\bibitem{shane} http://www.srl.caltech.edu/\~\,shane/sensitivty.



\bibitem{Kolb}  E.W. Kolb and M. S. Turner,
{\it The Early Universe} (Addison-Wesley, Reading, MA, 1990).

\bibitem{Kampfer} B. Kampfer, Ann.Phys. {\bf 9}, 605 (2000).

\bibitem{schwarz2} D.J. Schwarz,  Ann.Phys. {\bf 12}, 220 (2003);
                   D. Boyanovsky, H.J. de Vega, and D.J. Schwarz,
                    Ann. Rev. Nucl. Part. Sci. {\bf 56}, 441, (2006).



\end{thebibliography}
\end{document}